\documentclass[12pt]{article}
\usepackage[dvips]{graphicx}
\usepackage{pdproc}
\usepackage{psfrag}

  \makeatletter 
  \def\@cite#1{[#1]} 
  \makeatother    
\begin{document}

\renewcommand{\thefootnote}{\alph{footnote}}
\newcommand{\bbms}{$B_s\!-\!\ov{B}{}_s\,$\ mixing}

\newcommand{\lt}{\left}
\newcommand{\rt}{\right}
\newcommand{\no}{\nonumber}
\newcommand{\nn}{\nonumber \\}
\newcommand{\ov}[1]{\overline{#1}}
\newcommand{\eq}[1]{(\ref{#1})}
\newcommand{\be}{\begin{equation}}
\newcommand{\ee}{\end{equation}}
\newcommand{\bea}{\begin{eqnarray}}
\newcommand{\eea}{\end{eqnarray}}
\newcommand{\cL}{{\cal L}}
\newcommand{\cO}{{\cal O}}
\newcommand{\tg}{{\tilde g}}
\newcommand{\eff}{{\mbox{\rm \scriptsize eff}}}
\newcommand{\mtpole}{{m_t^{\mbox{\rm \scriptsize pole}}}}
\newcommand{\tq}{{\tilde q}}
\newcommand{\td}{{\tilde d}}
\newcommand{\tb}{{\tilde b}}
\newcommand{\ts}{{\tilde s}}
\newcommand{\tl}{{\tilde l}}
\newcommand{\te}{{\tilde e}}
\newcommand{\stp}{{\tilde t}}
\newcommand{\gev}{\ensuremath{\,\mbox{GeV}}}
\newcommand{\mgut}{{M_{\mbox{\rm \scriptsize GUT}}}}
\newcommand{\mten}{{M_{10}}}
\newcommand{\mpl}{{M_{\mbox{\rm \scriptsize Pl}}}}
\newcommand{\bra}[1]{\langle \, #1 \, | }
\newcommand{\ket}[1]{| \, #1 \, \rangle }
\newcommand{\diag}{{\mathrm{diag}}}

{\normalsize\sf\hspace*{-0.8cm}
PITHA-04/15
\\
FERMILAB-CONF-04-296-T
}
\vspace{0.7cm}

\title{
 Hadronic vs leptonic flavor and CP violation in SUSY SO(10){\begingroup%
\def\thefootnote{\fnsymbol{footnote}}\footnote[1]{
Talk given at SUSY 2004, June 17-23, 2004, Tsukuba, Japan.
}\endgroup}
}

\author{SEBASTIAN J\"AGER\footnote{Speaker.}\footnote{
Supported in part by
the DFG SFB/TR 9 ``Computergest{\"u}tzte
Theoretische Teilchenphysik''.}$^{\mathrm c}$
 and ULRICH NIERSTE$^{\mathrm d}$
}

\address{ 
$^{\mathrm c}$Institut f\"ur Theoretische Physik E, 
RWTH Aachen, 52056 Aachen, Germany \\
$^{\mathrm d}$Fermi National Accelerator Laboratory, Batavia,
        IL 60510-500, USA
}

\abstract{
We study hadronic and leptonic flavor physics in a SUSY SO(10) model
proposed by Chang, Masiero, and Murayama, which links $b\to s$ transitions to
the observed large atmospheric neutrino mixing angle.
We find large effects in \bbms\ and $BR(\tau\to\mu\gamma)$
and comment on $B_d\to\phi K_S$.
}

\normalsize\baselineskip=15pt

\section{Introduction}
\label{sec:intro}
Within SUSY GUTs, quarks and leptons are unified into irreducible symmetry
multiplets, opening the possibility of links between hadronic and
leptonic flavor structures. This raises the question if there exist
scenarios where the large atmospheric and solar mixing
angles can manifest themselves in the hadronic sector.
On the other hand, due to the presence of
scalar quarks and leptons, there are many new parameters that can violate
flavor in supersymmetric
theories. Radiative corrections to these above
the GUT scale imply mass differences
and flavor violation even for universal soft terms at the Planck scale,
and certain mixing angles that would otherwise be unphysical
can be rendered observable~\cite{Barbieri:1994pv}.
Chang, Masiero, and Murayama (CMM)
have proposed an SO(10) model in which the
large $\nu_\mu$--$\nu_\tau$ mixing angle can affect transitions
between right-handed $b$ and $s$ quarks~\cite{cmm}.
 The SO(10)-symmetric superpotential has the form
\be
W_{10} = \frac{1}{2} \mathbf{16}^T Y^U \mathbf{16} \, \mathbf{10}_\mathbf{H} 
  \, + \, 
   \frac{1}{\mpl} \frac{1}{2} \mathbf{16}^T \tilde{Y}^D  \mathbf{16}
     \, \mathbf{10_H^\prime} \mathbf{45_H} 
 + \frac{1}{\mpl} \frac{1}{2}  \mathbf{16}^T Y^M \mathbf{16}  
   \, \mathbf{\ov{16}_H} \mathbf{\ov{16}_H}. \label{w10}
\ee
Here $\mathbf{16}$ is the usual spinor comprising the matter
superfields and the other fields are Higgs superfields in the
indicated representations. $Y^U$ is a symmetric
$3\times 3$ matrix in generation space containing the large top Yukawa
coupling. The two dimension-5 terms involve the Planck mass $\mpl$
and further Higgs fields in the indicated representations.
At some scale $\mten$ between the Planck and GUT scales 
the $\mathbf{45_H}$ and $\mathbf{\ov{16}_H}$ acquire VEVs
$v_{45}$ and $v_{\ov{16}}$, and SO(10)
is broken to SU(5), which is broken to the MSSM at the GUT scale.
The SU(5) superpotential reads
\be
W_5 = \frac{1}{2} \Psi^T Y^U \Psi \mathbf{5}_\mathbf{H} \, + \,  
        \Psi^T Y^D \Phi \, \mathbf{\ov{5}_H} \, +\, 
        \Phi^T  Y^\nu N \, \mathbf{5}_\mathbf{H} 
+\, 
         \frac{1}{2}\frac{v_{\ov{16}}^2}{\mpl} \, 
         N^T Y^M N , \label{w5} 
\ee
with $Y^D\propto \tilde Y^D v_{45}/\mpl$.
The last term in \eq{w5} generates small neutrino masses via the standard
seesaw mechanism.  From $m_t \gg m_c$ we observe a large hierarchy
in $Y^U \approx Y^\nu$, which must be largely compensated in $Y^M$ in order to
explain the observed pattern of neutrino masses. This is achieved in a
natural way by invoking flavor symmetries, the simplest of which
render $Y^U$ and $Y^M$ simultaneously diagonal. This defines the
$(U)$-basis. In this basis the remaining Yukawa matrix
$Y^D \approx {Y^E}^T$, responsible for the masses of down-type
quarks and charged leptons, has the form
$Y^D = V_{\rm CKM}^*\,\diag(y_d, y_s, y_b)\,U_{\rm PMNS}$.
Here $V_{\rm CKM}$ and $U_{\rm PMNS}$
encode flavor mixing in the quark and lepton sectors, and
certain diagonal phase matrices have been omitted. 
The nonsymmetric structure of $Y^D$ is possible because the
corresponding dimension-5 term in \eq{w10} transforms reducibly under
SO(10). Note also that generically $\tan\beta = \cO(\mten/\mpl)$.
The matter supermultiplets $\Psi$, $\Phi$ and $N$ are the usual
$\mathbf{10}$, $\mathbf{\ov 5}$, and $\mathbf{1}$ from the
decomposition of the $\mathbf{16}$. We also have
$\Psi \supset (Q, U, E)$, $\Phi \supset (D, L)$,
$\mathbf{10}_\mathbf{H} \supset \mathbf{5}_\mathbf{H} \supset H_u$, and
$\mathbf{10_H^\prime} \supset \mathbf{\ov{5}_H} \supset H_d$.

The soft SUSY-breaking terms are assumed universal near the Planck
scale. The large Yukawa coupling in $Y^U$ now renormalizes the sfermion
mass matrix, keeping it diagonal in the $(U)$-basis
but splitting the mass of the third from those of the first two generations
for each MSSM sfermion multiplet.
The diagonalization of $Y^D$  involves the rotation of $\Phi$ in \eq{w5}
with $U_{\rm PMNS}$.
Since  $\Phi$ unifies left-handed (s)leptons with right-handed
down-type (s)quarks, the large atmospheric mixing angle will appear in 
the mixing of ${\tilde b}_R$ and ${\tilde s}_R$.

\section{RG analysis of the CMM model}
The large Yukawa coupling $y_t$ driving all nonuniversal
renormalization-group effects, its own behavior under RG evolution
is crucial. In the MSSM and the considered GUTs,
$y_t$ possesses an IR quasi-fixed point, and for sufficiently
small values of $\tan\beta$ the low-energy value of $y_t$ will reside
above the fixed-point trajectory. For the situation in the CMM model, see
\begin{figure}[htb]
\begin{center}
  \psfragscanon
        \psfrag{LogMuGev}{$\log \mu [\mbox{GeV}]$}
        \psfrag{yUt}{$y_t$}
        \psfrag{tanbexFFFFF}{$\tan\beta=3.00$}
        \psfrag{tanbcrFFFFF}{$\tan\beta=2.30$}
        \psfrag{tanbcrm04FF}{$\tan\beta=1.90$}
        \resizebox{80mm}{!}{\includegraphics{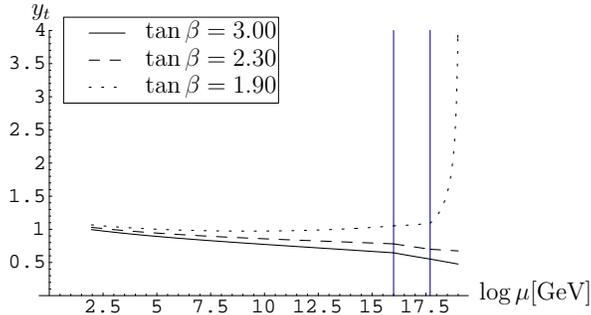}}
\caption{RG evolution of $y_t$. See text for explanation.
\label{fig:ytrun}}
\end{center}
\end{figure}
Fig.~\ref{fig:ytrun}. The two vertical lines indicate the GUT and
SO(10)-breaking scales. The dashed line is the ``critical'' trajectory,
corresponding to the SO(10) fixed point of $y_t/g$. The dotted and solid lines
are examples corresponding to $y_t(\mu)$ for $m_t=174$ GeV and the
stated values of $\tan\beta$, with mild dependence on other supersymmetric
parameters (a shift in $m_t$ can be compensated by a shift in
$\tan\beta$). In general, $y_t$ becomes nonperturbative at high energies
if it resides above the fixed point (dotted line). Perturbativity
being essential for the model to be predictive,
we restrict ourselves to $y_t$ below the critical line.

As anticipated in the introduction, in the $(U)$-basis
the mass matrices
of the right-handed down-type squarks and of the left-handed sleptons,
renormalized at the weak scale, have the form
$
        m^{2,(U)}_{\td} = \diag \left(m^2_{\td_R},m^2_{\td_R},
                                 m^2_{\td_R} - \Delta_\td\right)$,
\ \ 
$        m^{2,(U)}_{\tl} = \diag \left(m^2_{\tl_1}, m^2_{\tl_1},
                                 m^2_{\tl_1} - \Delta_\tl\right) ,
$
with $\Delta_\td \approx \Delta_\tl$ due to SU(5) and SO(10) GUT relations.
This nonuniversal structure then induces flavor-changing couplings of
gluinos and neutralinos once the charged-lepton and down-type-quark
Yukawa matrices
are diagonalized, both of which originate from $Y^D$ in~(\ref{w5}) and
are the transpose of each other:
$Y^E = U_{\rm PMNS}^T \hat Y^E U_E$,\ \ \ 
$Y^D = V_{\rm CKM}^* \hat Y^D U_D$,
$Y^D \approx {Y^E}^T$ .
Specifically, the couplings of right-handed
down-type squarks to gluinos and quarks now contain an element
of~$U_D \approx U_{\rm PMNS} \equiv U$, as do the couplings of
left-handed sleptons to neutralinos.
Imposing (approximate) Yukawa unification only for the bottom and tau,
one is left with the weaker statement
$
        |{U_D}_{23}| \approx |{U_D}_{33}| \approx |U_{\mu 3}|
        \approx |U_{\tau 3}| ,
$
suggesting large FCNC involving second-to-third-generation
transitions.

Besides $\tan\beta$, we choose as low-energy input parameters
the approximately universal first-generation squark mass $m_\tq$,
the gluino mass $m_\tg$,
and the down-squark $A$-parameter $a_d$,
all of which are constrained by phenomenology.
We use the MSSM and GUT RGEs to relate them
to Planck-scale parameters and compute the low-energy quantities
$\Delta_\td$ and $\Delta_\tl$ from the RGE solutions, with the dominant
contributions coming from above $\mten$.

\section{\bbms, $\tau\to\mu\gamma$, and $B_d\to\phi K_S$}
\begin{figure}[htb]
\begin{center}
\psfragscanon
\psfrag{mgl195}{$m_{\tg}=195$ GeV}
\psfrag{msq}{$m_\tq$}
\psfrag{adbymsq}[Bl][Bl][1][180]{$a_{d}/m_\tq$}
        \resizebox{70mm}{!}{\includegraphics{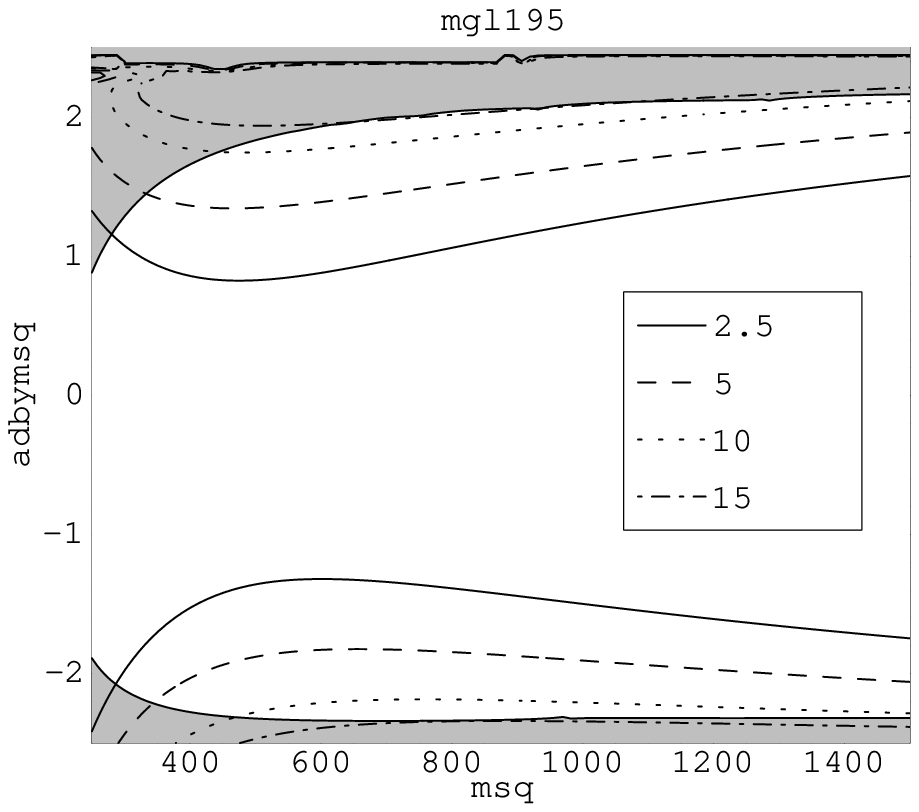}}
\hspace{7mm}
\psfrag{m6}{$10^{-6}$}
\psfrag{mexp}{$6 \cdot 10^{-7}$}
\psfrag{m6p5}{$10^{-6.5}$}
\psfrag{m7}{$10^{-7}$}
\psfrag{m7p5}{$10^{-7.5}$}
\psfrag{m8}{$10^{-8}$}
\psfrag{m9}{$10^{-9}$}
\psfrag{mgl}{$m_{\tg_3}$}
\psfrag{msq}{$m_\tq$}
\psfrag{adbymsq}[Bl][Bl][1][180]{$a_{d}/m_\tq$}
\psfrag{mcut}{$m_{{\tilde \tau}_R}^{\mbox{\scriptsize min}}$}
        \resizebox{67mm}{!}{\includegraphics{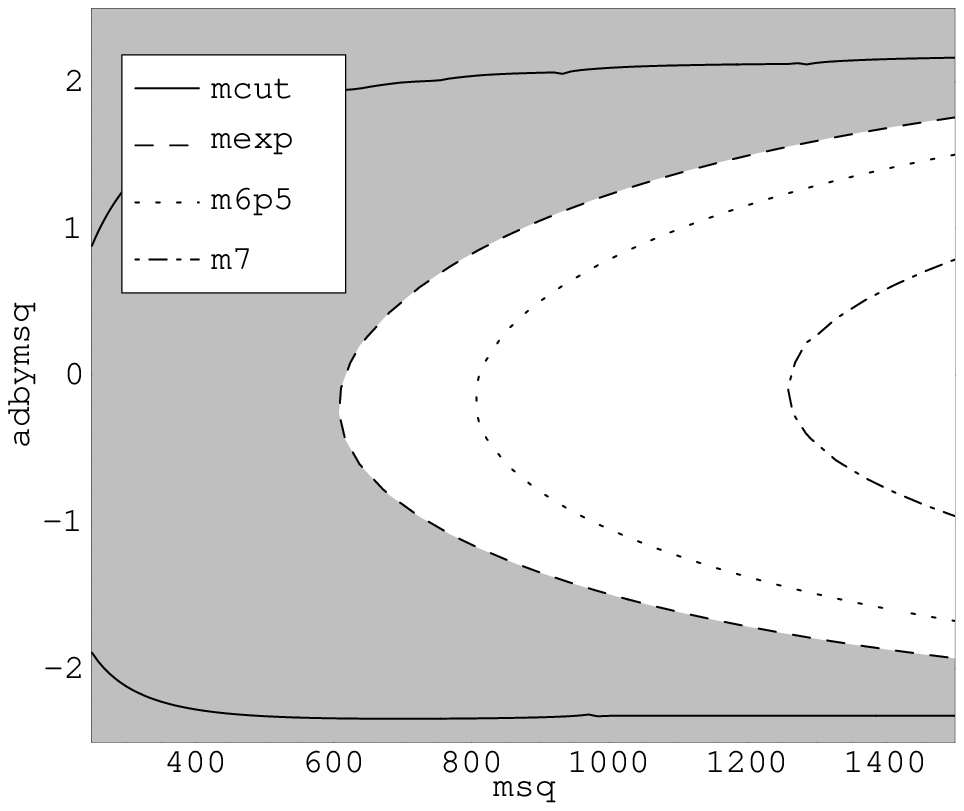}}
\caption{Left: New contribution to \bbms. Right: $BR(\tau\to\mu\gamma)$.
For explanation, see text.
\label{fig:bb}}
\end{center}
\end{figure}

The dominant new contributions to \bbms\ in the CMM model are $\cO(\alpha_s^2)$
corrections from one-loop box 
diagrams with gluinos and squarks. One obtains the mixing amplitude
\be
 M_{12} = \frac{G_F^2\, M_W^2 }{32  \pi^2 \, M_{B_s}} \, \lambda_t^2
        \, 
          \lt( C_L + C_R \rt) \bra{B_s} O_L
        \ket{\ov{B}_s} . \label{eq:m12}
\ee
Here $ \bra{B_s} O_L \ket{\ov{B}_s}$ (with $\bra{B_s} B_s \rangle = 2
E V$) is the matrix element of the
usual Standard Model four-quark effective operator
$O_L = \bar s_L \gamma_{\mu} b_L \; \bar s_L \gamma^{\mu} b_L$ and
$C_L$ is due to Standard Model $W-t$ exchange. Note
that in a general model several operators with independent hadronic
matrix elements arise, while we encounter only
gluino-squark boxes generating the parity reflection of $O_L$, with
the Wilson coefficient
\be
        C_R = 
        \frac{\Lambda_3^2}{\lambda_t^2} 
        \frac{8 \pi^2 \alpha_s^2 (m_{\tilde g}) }{G_F^2 M_W^2 m_{\tilde g}^2 }
        \left[ \frac{\alpha_s(m_{\tilde g})}{\alpha_s(m_b)}\right]^{6/23}
        S^{(\tg)} .  \label{eq:CR}
\ee
Furthermore, $\lambda_t= V_{ts}^* V_{tb}$ is the applicable Standard
Model flavor-mixing parameter and
$
        |\Lambda_3| = |U_{\mu 3}| |U_{\tau 3}| \approx \frac{1}{2}
$
is the relevant combination of mixing-matrix elements in the right-handed
sdown sector, and $S^{(\tg)}$ is a dimensionless function of the
squark and gluino masses. Note the twofold enhancement of $C_R$ due
to the large atmospheric mixing and the large strong coupling
constant. This is, however, partially offset by a smaller loop function
$S^{(\tg)}$. The neutral $B_s$-meson mass difference is given
by $\Delta M_{B_s} = 2 | M_{12} |$,
while the phase of $M_{12}$ is responsible for mixing-induced CP
violation. Fig.~\ref{fig:bb} (left) shows a contour plot of the modulus
of the CMM contribution normalized to the Standard Model prediction in
the $(m_\tq,a_d)$ plane for $m_\tg=195$~GeV.
The shaded area in the plot is excluded by experimental constraints on the
sparticle spectrum.
We observe that the Standard Model prediction of about $17.2\,\mathrm{ps}^{-1}$
can be exceeded by a factor of 16. The effect decreases rapidly
with increasing gluino mass.
Note that the phase of $\Lambda_3$ is undetermined, so that
there is potentially large CP violation in decays such as
 $B_s \to \psi \phi$ and $B_S\to\psi \eta^{(\prime)}$.

We have also computed the amplitudes
for $\tau\to\mu\gamma$ and $B_d\to\phi K_S$.
The former now also depends on the $\mu$-parameter,
but it is clear that rates are in general large compared to the GIM-suppressed
Standard Model case. The right plot in Fig.~\ref{fig:bb} shows contours
of constant $BR(\tau\to\mu\gamma)$, with the shaded area excluded by
the (old) Belle upper bound, demonstrating that this mode constrains
the CMM model.
$B_d\to\phi K_S$ is more involved
because several operators contribute. Using QCD factorization, we find
the chromomagnetic penguin operator to be associated with the dominant
SUSY contribution, which could give a $\cO(1)$ correction to
the Standard Model amplitude, again with an unconstrained phase. Whether
this can explain a large deviation of the time-dependent CP asymmetry from
$\sin(2\beta)$ when reconciled with the experimental constraints
on $BR(\tau\to\mu\gamma)$ remains to be assessed.

In conclusion, we find that there are GUT models connecting
the atmospheric mixing angle with hadronic observables and being predictive
for flavor physics. We find large effects in several
observables. Our analysis is complementary to other works studying more general
setups~\cite{gutflav}, whereas we examine a more predictive scenario
quantitatively.

\bibliographystyle{plain}

\end{document}